\documentclass{article}

\usepackage{arxiv}

\usepackage[utf8]{inputenc} % allow utf-8 input
\usepackage[T1]{fontenc}    % use 8-bit T1 fonts
\usepackage{hyperref}       % hyperlinks
\usepackage{url}            % simple URL typesetting
\usepackage{booktabs}       % professional-quality tables
\usepackage{amsfonts}       % blackboard math symbols
\usepackage{nicefrac}       % compact symbols for 1/2, etc.
\usepackage{microtype}      % microtypography
\usepackage{lipsum}		% Can be removed after putting your text content
\usepackage{graphicx}
\usepackage[numbers]{natbib}
\usepackage{doi}
\usepackage{amsmath}

\title{Integrating state-sequence analysis to uncover dynamic drug-utilization patterns to profile heart failure patients}

\author{ \href{https://orcid.org/0009-0007-4416-19210}{\includegraphics[scale=0.06]{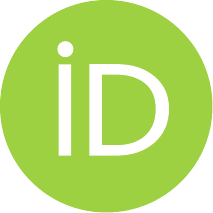}\hspace{1mm}Nicole ~Fontana}\thanks{Use footnote for providing further
		information about author (webpage, alternative
		address)---\emph{not} for acknowledging funding agencies.} \\
	MOX, Department of Mathematics,\\
	HDS, Health Data Science Center, Human Technopole, Milan, Italy\\
	\texttt{nicole.fontana@polimi.it} \\
	\And
	\href{https://orcid.org/0000-0001-9030-7011}{\includegraphics[scale=0.06]{orcid.pdf}\hspace{1mm} Laura~Savaré} \\
	MOX, Department of Mathematics,\\
	HDS, Health Data Science Center, Human Technopole, Milan, Italy\\
	HRP, National Centre for Healthcare Research \& Pharmacoepidemiology, University of Milano-Bicocca, Milan, Italy \\
	\texttt{laura.savare@polimi.it} \\
 	\And
	\href{https://orcid.org/0000-0003-0165-1983}{\includegraphics[scale=0.06]{orcid.pdf}\hspace{1mm} Francesca~Ieva} \\
	MOX, Department of Mathematics,\\
	HDS, Health Data Science Center, Human Technopole, Milan, Italy\\
	\texttt{francesca.ievae@polimi.it} \\
}

\renewcommand{\shorttitle}{\textit{arXiv} Template}

\hypersetup{
pdftitle={A template for the arxiv style},
pdfsubject={q-bio.NC, q-bio.QM},
pdfauthor={David S.~Hippocampus, Elias D.~Striatum},
pdfkeywords={First keyword, Second keyword, More},
}

\begin{document}
\maketitle

\begin{abstract}
Globally, the incidence of heart failure is increasing, and its principal treatment involves drug therapy. However, widespread non-adherence to therapies is prevalent among heart failure patients and often results in worsening health conditions and an increase in hospital admissions. This study aims to develop an innovative approach, the State-Sequence analysis, to profile heart failure patients based on different drug-utilization patterns. These patterns aim to capture both the multidimensional and dynamic effects of therapies. Subsequently, the study explores how combining clustering algorithms with this technique influences overall patient survival. Findings highlight the importance of continued drug therapy after the first hospitalization in improving heart failure prognosis, irrespective of its severity. The proposed approach can assist healthcare specialists in evaluating the pathways provided to patients, allowing for a change in analysis from a transversal and syntactical approach to a holistic one that leverages statistical tools that are slightly more complex than traditional methods. Moreover, because of the many options available for defining states, temporal granularity, and spacing metrics, SSA is a flexible method applicable to various epidemiological scenarios.
\end{abstract}

% keywords can be removed

\section{Introduction}\label{sec:1}
Over the past decade, heart failure (HF) has emerged as a significant medical concern~\cite{cite_intro_1}, with data indicating that the persistent efforts to treat and manage HF have yet to reduce the burdens of mortality and hospitalization substantially~\cite{cite_intro_2}. The global prevalence of heart failure is increasing, attributed to the growing incidence of an ageing population~\cite{cite_1}. The fundamental objectives of heart failure treatments encompass reducing mortality, preventing recurrent hospitalization, and enhancing both clinical status and functional capacity~\cite{cite_15}.
The basis of its treatment primarily involves pharmacotherapy. The most commonly prescribed drugs in the guidelines of 2005~\cite{cite_27} are angiotensin-converting enzyme inhibitors (ACEi) or angiotensin-receptor blockers (ARBs), collectively referred to as renin-angiotensin system inhibitors (RAS). Additionally, beta-blockers (BB) and anti-aldosterone agents (AA) play crucial roles in this therapeutic approach~\cite{cite_13, cite_14}. Guidelines support the combined use of these therapies, as studies have demonstrated their effectiveness in reducing mortality and morbidity among heart failure patients with reduced ejection fraction~\cite{cite_15}.
Numerous studies underscore high mortality and re-hospitalization rates among heart failure patients~\cite{cite_16, cite_17}. These elevated rates can be attributed to the phenomenon of non-adherence to medications~\cite{cite_18, cite_19}, a prevalent issue in HF patients with estimated adherence rates ranging from 10$\%$ to 93$\%$~\cite{cite_20, cite_21, cite_3}.
Moreover, treatments have been improved over time due to advances in understanding the pathophysiology of heart failure~\cite{cite_28}, underscoring the necessity of a systematic approach to assessing polytherapy adherence that can consider expanding pharmacological treatment options over time.

%Limitation of existing methods 
Examining adherence to combined therapy is crucial. To achieve this, it is essential to employ a tool capable of accurately defining adherence to various medications.
While recent years have seen the introduction of methodologies to address this information~\cite{cite_22, cite_23, cite_24}, they might not be appropriate for dynamically evaluating complex longitudinal care patterns. Simultaneously, capturing the effects of dynamic changes in drug adherence on patients' outcomes is decisive, as drug intake and adherence patterns can evolve throughout therapy. 
%our proposal
For this reason, we focus on the therapeutic pathways administered to HF patients using an innovative method, at the expense of more traditional ones, to have a deeper description capable of evaluating the temporal order of drug use to extract drug-utilization patterns and their association with health outcomes.
We propose using State Sequence analysis (SSA), an emerging technique in social sciences that has gained interest in epidemiology, to represent care consumption. This technique provides valuable insights into the intervals, the timing of treatment patterns, and the effectiveness of various treatment sequences~\cite{cite_25, cite_26}.
%data
We apply such methods to an administrative database of hospitalized HF patients in the Lombardy region from 2006 to 2012. The use of this data is motivated by their capacity to provide extensive and systematically collected heterogeneous data, allowing them to effectively capture complex care trajectories within real-world healthcare settings~\cite{cite_29, cite_7}.

\section{Data and cohort selection}
%2.1 Study setting and data sources
The data used in this study are taken from the Healthcare Utilization Databases of Lombardy, a Region of Italy with approximately 16\% of its population (more than 10 million residents)~\cite{cite_8}.
This database collects patients hospitalized for heart failure between 2006 and 2012 and provides information about hospitalizations and drug purchases. 
Hospitalization records, coded according to the International Classification of Diseases, Ninth Revision Clinical Modification (ICD-CM-9) classification system, offer details on the primary diagnosis, co-existing conditions, and procedures~\cite{cite_21_LMM}. In drug purchase records, the Anatomical Therapeutic Chemical (ATC) classification system ~\cite{cite_22_LMM} is utilized to categorize each drug, and the Defined Daily Dose (DDD) metric ~\cite{cite_23_LMM} is employed to extract information regarding the number of treatment days covered by a specific drug.

%2.2 Study population
A total cohort of 187499 patients with a principal diagnostic code of HF was initially identified. Patients accumulated person-years from the date of hospital discharge, defined as the index date, until the death, censorship due to emigration to another region or state, or the study's endpoint (31-12-2012). The study period was split into two parts: the observation period (365 days from the index date) and the follow-up period (started 365 days after the index date). 

The cohort selection is then determined based on specific criteria: patients who died during the observation period were excluded. A 5-year wash-out period from 2000 to 2005 was used to consider only new incident HF patients (patients with no contact with the health care system in the previous five years due to HF). Looking at the observation period, any individual with a censoring date within this period and without any pharmacological purchase related to disease drugs was excluded from the study.  
Finally, due to computational issues arising from the complexity of the methods, we decided to apply stratified random sampling for the outcome of interest, that is, mortality. Thus, a total of 35842 patients met the study selection criteria. Cohort participants were evaluated for several covariates. Baseline variables included age, sex, comorbidities, and procedures.
In addition, the Multisource Comorbidity Score (MCS), a comorbidity index derived from inpatient diagnostic information and outpatient treatment data, was computed for each member of the cohort~\cite{cite_9}. Patients were then classified based on their clinical profiles: low (0-4), intermediate (5-9), and high ($\geq$ 10).

\label{sec:2}

\section{Methodology: State sequence analysis}
State-sequence analysis was born to understand how events in ordered sequences can be related to the outcome of interest. It is a well-defined technique in sociology that assesses how the chronological order of events in subgroups can lead to different social behaviours~\cite{cite_10}. The main objective of sequence methods is to extract simplified, workable information from longitudinal data. Combined with cluster analysis, this method allows us to identify sequence descriptors that can be used in predictive models. 

\subsection{Sequences definition}\label{3.1}
Sequences are mathematical objects that contain elaborated information coming from the original data. A state sequence is formally defined as an ordered list of elements selected from a finite set. These elements can precisely denote a particular state (for example, treatment coverage), with their positions representing relative time points and carrying specific interpretations. The sequence attributed to each patient is regarded as a conceptual unit.
State-sequences are characterized by two properties: an \textit{alphabet}, which is the list of all possible elements, and a \textit{time axis}, which assigns to each state (i.e., the unit of measure) an element~\cite{cite_11}. 

This definition enables assigning specific sequences to each patient, one for each drug of interest, describing their coverage patterns over time. We addressed the multichannel sequence analysis (MSA) using the extended alphabet (EA) approach to deal with combined therapies.
In particular, the elements of each channel are combined to build a single set of super-elements, called an extended alphabet, each super-element being defined by combining one element from each of the original channels~\cite{cite_30}. 
In general, according to~\cite{cite_13_TESI}, it is possible to divide the SSA into three main steps: (i) identification sequences elements; (ii) measuring sequence dissimilarity; (iii) clustering sequences.

\subsection{Dissimiliarity measures}\label{3.2}
After analyzing the sequences, it is necessary to define a metric to measure their dissimilarity. 
A dissimilarity measure is defined as a quantitative evaluation of the level of mismatch between two sequences~\cite{cite_10}. This level of mismatch gives an idea of the (lack of) resemblance between the sequences and, therefore, helps compare and cluster sequences.
The sequences may encompass various interdependent dimensions, including elements experienced, distribution, timing, duration, and sequencing. Consequently, the similarity between two sequences can be measured by emphasizing one or multiple dimensions. Thus, comparing sequences poses a significant challenge in SSA, potentially leading to different partitions depending on the adopted metric.
Dissimilarity metrics are classified into two main classes: edit-based distances and counts of common attributes-based distances~\cite{cite_44_TESI}. Edit-based metrics measure the distance between two sequences by counting the minimum number of (weighted) edit operations required to turn one sequence into a perfect copy of the other. On the other hand, attributes-based metrics measure the distance between sequences by counting the number of (weighted) common attributes.
In our analysis, we opted for the widely used edit-based metric, Optimal Matching (OM) distance. This metric calculates edit distances as the minimal cost, considering insertion, deletion, and substitution operations required to convert one sequence into another.
The rationale behind this choice is driven by the parameterization of operation costs, providing flexibility and allowing for a trade-off between time and contemporaneity of the states~\cite{cite_12}. Insertion and deletion operations preserve the contemporaneity of the states but may distort time, while substitution operations do the opposite, conserving time but altering the ordering of states. Unlike other metrics, OM does not conduct pairwise comparisons but considers similar shifted patterns. In our context, this feature is crucial, as we aim to bring two patients closer if they exhibit similar behaviour in taking therapies, irrespective of differences in the duration of intake for a brief period.

\subsection{Clustering sequences}\label{3.3}
Once the dissimilarity matrix is computed, a cluster analysis can be performed to construct partitions of the sequences into distinct groups.
Two clustering algorithms are applied to our work: hierarchical clustering and partitioning around medoids (PAM).
To evaluate the partition obtained, we use three different metrics to assess their capability: 
\begin{itemize}
    \item Point Biserial Correlation (PBC)~\cite{cite_PBC} measures the partition's capacity to reproduce the distance matrix.
    \item Hubert's C (HC) index~\cite{cite_HC} measures the gap between the partition obtained with the best partition that could have been obtained with this number of groups and this distance matrix.
    \item Average Silhouette Width (ASW)~\cite{cite_ASW} measures the coherence of observation assignments to a given group.
\end{itemize}
We obtain the final clustering by applying hierarchical clustering to the whole set of sequences and choosing the best number of clusters, maximizing the PBC and ASW, and minimizing the HC. 
Then, we initialized the PAM algorithm with the best result of the hierarchical clustering~\cite{cite_31}. Finally, we chose the partition that gives the best quality metrics and a significant interpretation of the clusters.

\begin{figure}[t]
\centering
    %\sidecaption[t]
    \includegraphics[scale=.35]{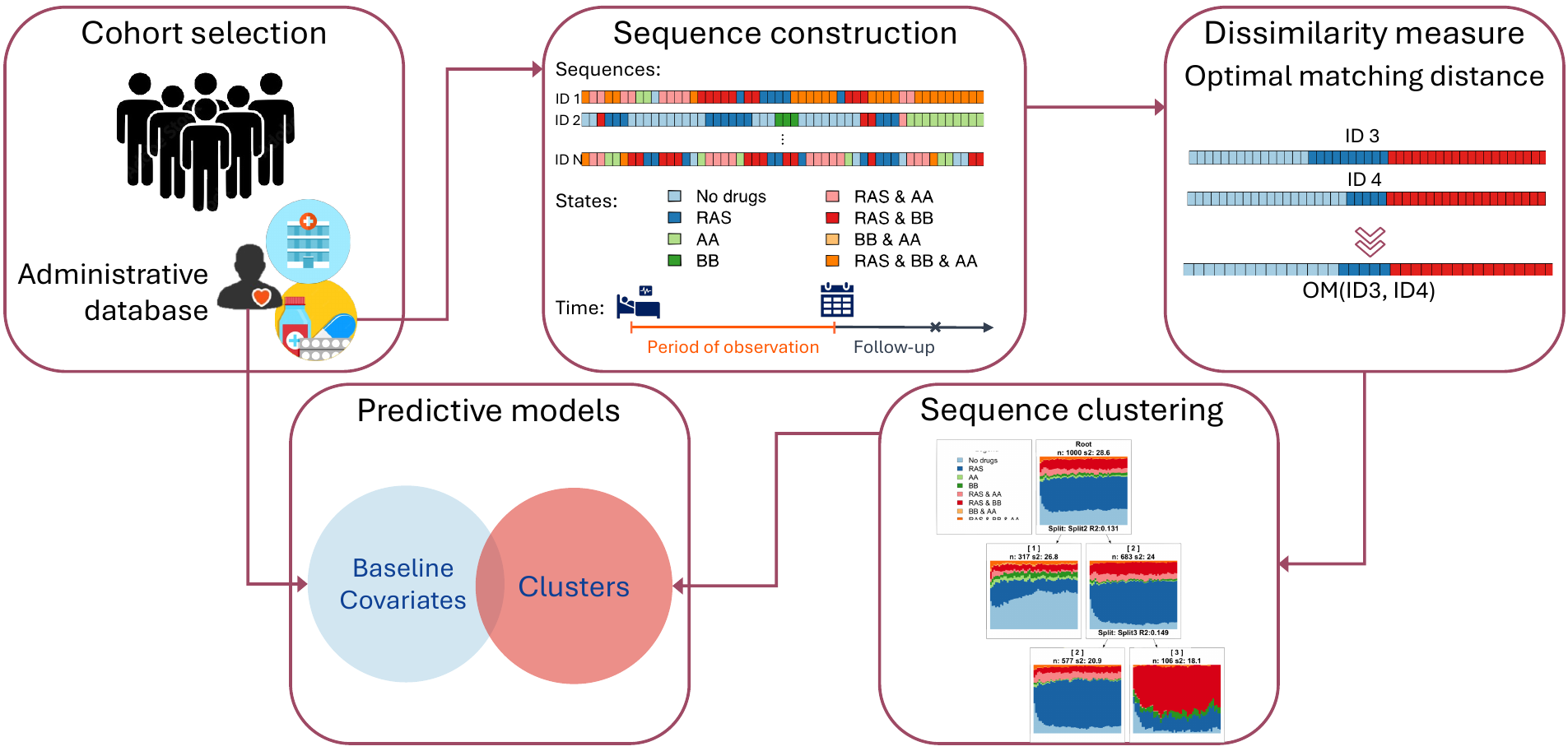}
    \caption{Algorithm pipeline of the SSA applied in this study.}
    \label{fig:pipeline}       
\end{figure}

Fig.~\ref{fig:pipeline} displays a schematic graphical representation of the entire methodological pipeline.

\label{sec:3}

\section{Results}
The analysis is developed as follows. State-sequence analysis (SSA) was initially utilized to investigate patients' behaviour regarding drug prescriptions. This involved capturing informative patterns that could influence patients' prognoses through sequence clustering. Following this, the clustering results incorporate sequence analysis into predictive models. This integration facilitates the assessment of the association between drug patterns and health-related outcomes.
\subsection{Combined sequence}
\begin{figure}[t]
%\sidecaption
\centering
\includegraphics[scale=0.75]{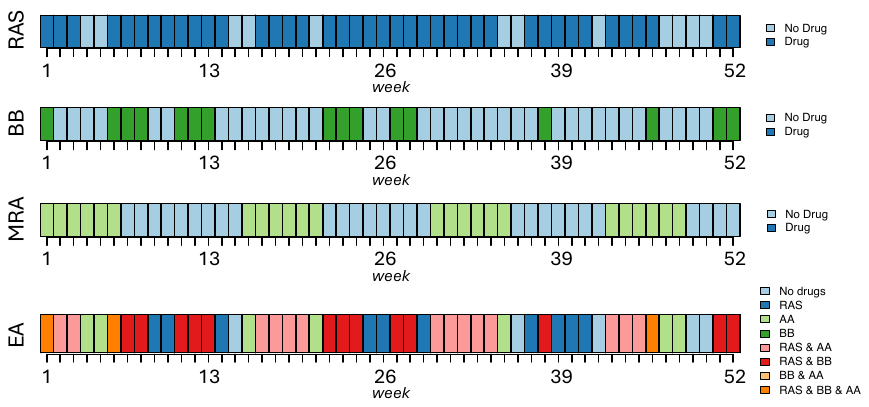}
\caption{Example of the computation of the combined sequence with texted alphabet (EA) for a patient starting from the single sequence of RAS, BB, and AA.}
\label{fig:seq_construction}       
\end{figure}

We investigate the prescription patterns of the RAS, BB, and AA drug classes for each patient during the observation period. This period allows us to monitor patient behaviours for a sufficiently extended duration to identify patterns, yet not excessively long to limit the impact of immortal bias~\cite{cite_31bis}.
To create the sequences, we retained all drug purchases made within the year of observation. If the drug coverage extended beyond that year but expired during the observation period, we truncated the coverage days to the last day of the observation period. We then eliminated any overlaps in purchases that occurred when the date of a drug purchase coincided with the coverage period of a previous purchase of the same drug type.
Each patient is associated with a sequence for the three drug classes. The sequences comprise 52 states, each representing a week of observation. Each state within a sequence is assigned a "No Drug" element if the drug purchase does not provide coverage for at least four out of seven days in the specific week and "Drug" otherwise.
For each patient, starting from the three single RAS, BB, and AA sequences, we build a combined sequence using the extended alphabet approach defined in Sect.~\ref{3.1} that combines the elements of the single sequences. This forms an alphabet consisting of eight elements, with each state in the alphabet representing the drugs taken by the patient in any given week. The entire procedure is shown in Fig.~\ref{fig:seq_construction}.

\begin{figure}[t]
%\sidecaption
\centering
\includegraphics[scale=0.55]{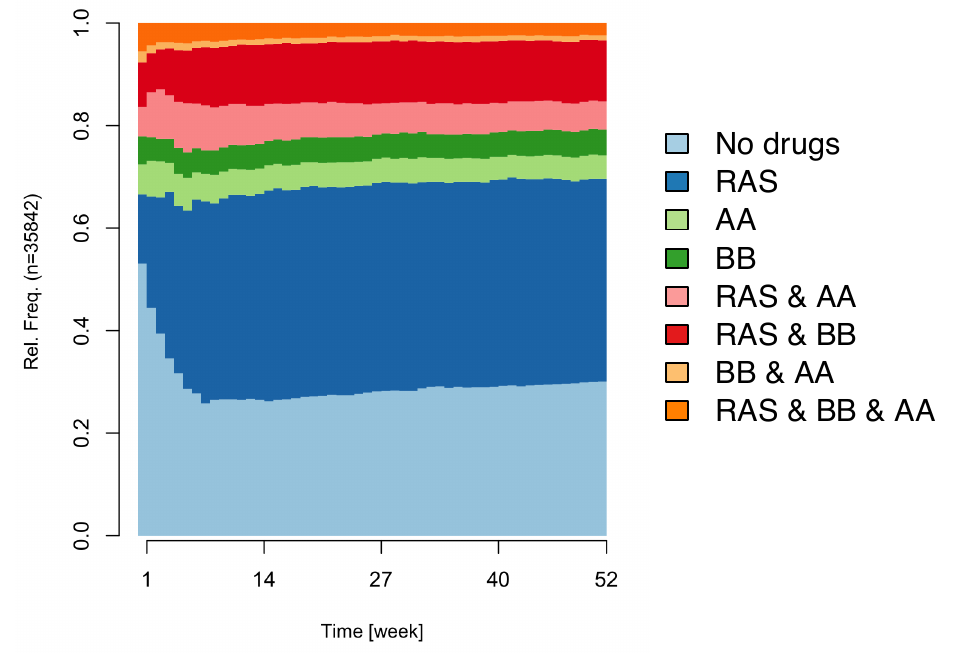}
\caption{State distribution plot of the combined sequences.}
\label{fig:seq_distribution}       
\end{figure}

\begin{figure}[t]
%\sidecaption
\centering
\includegraphics[scale=0.65]{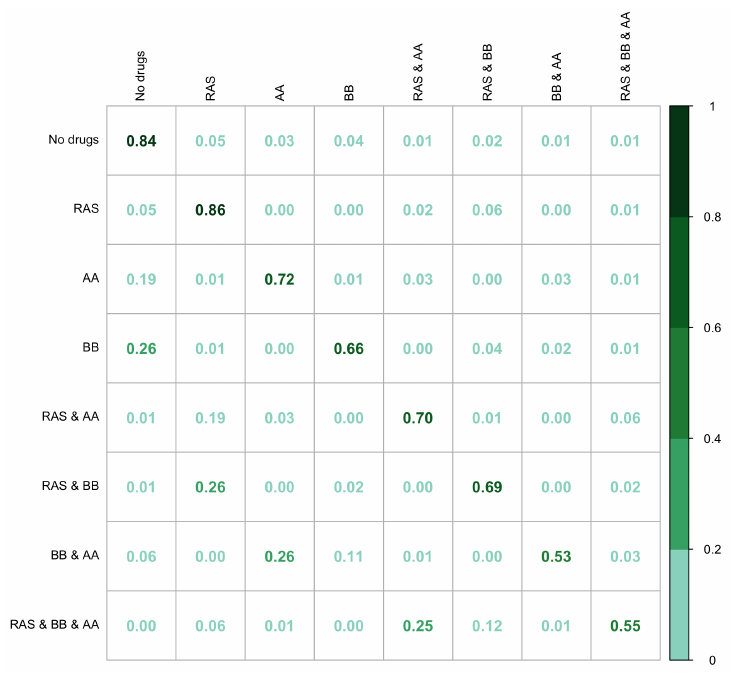}
\caption{Matrix of the transition rates between states of the combined sequence.}
\label{fig:trans_rates}       
\end{figure}

The analysis of the sequences performed through visual tools and statistical measures shows that over the first two months, the proportion of patients who do not take any therapy decreased in favour of a rise in the RAS assumption. 
A small fraction of patients take AA and BB separately, showing roughly the same proportion across the observation period. BB and AA are rarely taken by people who do not already take RAS. This information is displayed in Figure~\ref{fig:seq_distribution} through a state-distribution plot in which, for each state (i.e., each week), the frequency distribution of the elements is shown.
%state-distr plot
Looking at the transition rates, shown in Fig.~\ref{fig:trans_rates}, the probability of remaining in states that represent no treatments or monotherapy is elevated. 
However, BB and AA states are more unstable since the probability of stopping these therapies when accompanied by RAS is relatively high.
Generally, it is more likely for a patient to discontinue a therapy than to start or add a new one.
%transition rates

\subsection{Measuring dissimilarity and clustering sequences}
%Pairwise dissimilarities between extended alphabet sequences are computed as if they are a single channel.
Dissimilarities between combined sequences are calculated pairwise, treating them as a single channel. As motivated in Sect.~\ref{3.2}, the dissimilarity measure used for sequences comparison is based on the Optimal Matching distances, with indel costs set at one and substitution costs derived from transition rates between states.

\begin{figure}[t]
\centering
    %\sidecaption[t]
    \includegraphics[scale=.85]{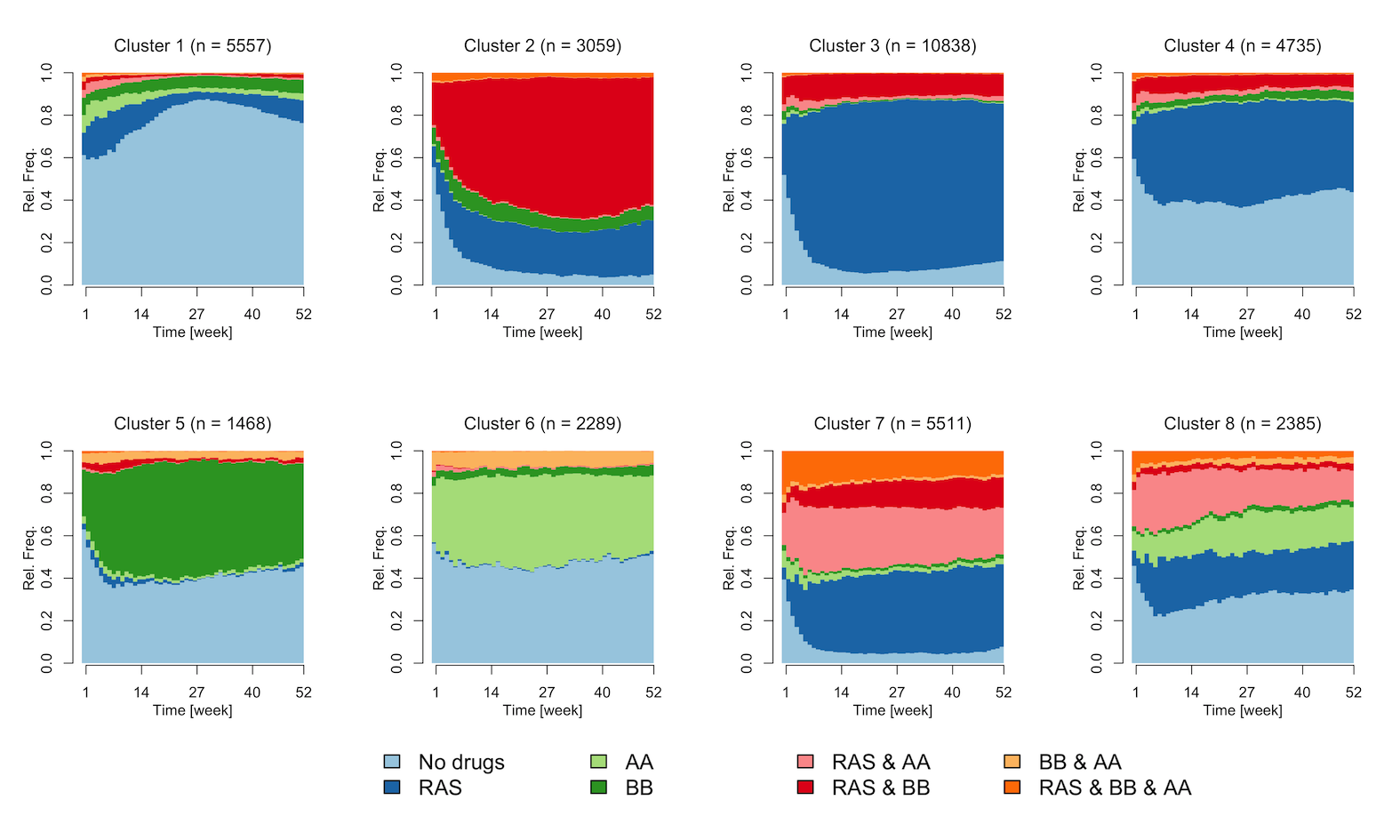}
    \caption{State distribution plot of the combined sequence by clustering partition results (hierarchical clustering with $k=8$).}
    \label{fig:cluster_comb}       
\end{figure}

Following the abovementioned clustering procedure, we derived an eight-cluster partition for the combined sequence through hierarchical clustering. This partition effectively captures the primary behavioural patterns of patients across the three classes of drugs, as illustrated in Fig.~\ref{fig:cluster_comb}. Clusters 1, 2, and 3 denote patients with constant therapy patterns for most of the observation period. Specifically, cluster 1 identifies patients as non-adopters of therapies, cluster 2 represents the combined intake of RAS and BB therapies, and cluster 3 indicates RAS as monotherapy. Patients falling into clusters 4, 5, and 6 are associated with half of those who took monotherapy (RAS, BB, AA, respectively), while the remaining individuals in these clusters were uncovered from any therapy throughout the entire year.
Lastly, Clusters 7 and 8 delineate heterogeneous behaviours concerning the three pharmacological classes.

\subsection{Including SSA-based representation into predictive models}
\begin{figure}[t]
\centering
    %\sidecaption[t]
    \includegraphics[scale=.06]{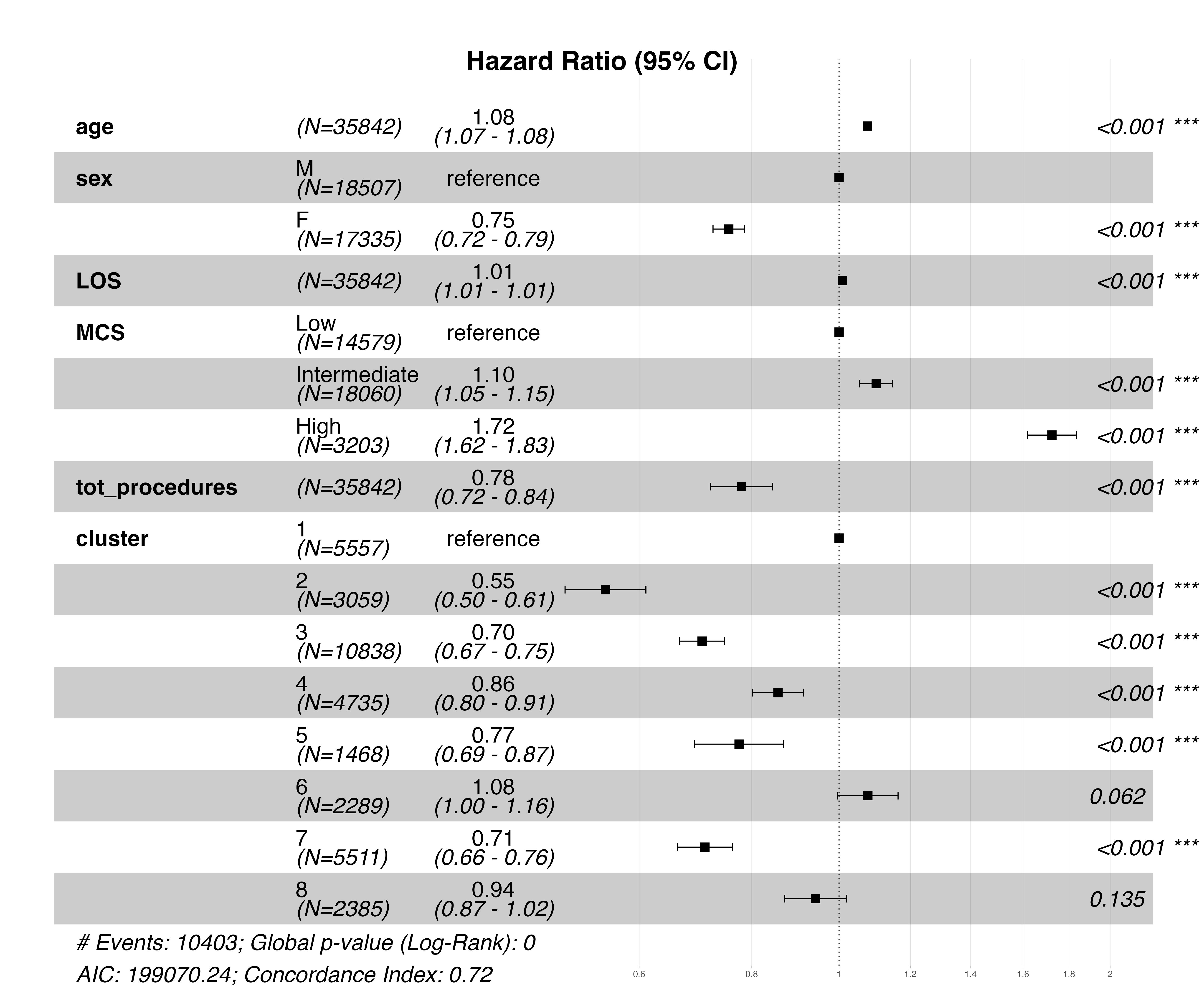}
    \caption{Hazard Ratio and its 95\% confidence interval of the survival model.}
    \label{fig:HR}       
\end{figure}

%HR e curve per I gruppi estremi
For each patient, we associated information with the respective cluster. Subsequently, we employed the Cox regression model to investigate whether and to what extent this information contributes to predicting overall patient survival. The model incorporates additional explanatory variables, including age, sex, multisource comorbidity score, and the total number of procedures the patients undergo.
Notably, as illustrated by the Hazard Ratio of the Cox model in Fig.~\ref{fig:HR}, all different care patterns (excluding clusters 6 and 8) contribute to an increased probability of survival with significance at 5\% compared to those characterized by non-adoption of therapies, emphasizing the protective nature of drug intake. Specifically, compared to the reference (cluster 1), patients in clusters 2, 3, and 7 exhibit a reduced probability of death by 45\% (CI, [39\% - 50\%]), 30\% (CI, [25\% - 33\%]), and 29\% (CI, [24\% - 34\%]), respectively. These clusters represent patients primarily utilizing RAS combined with BB, exclusively RAS, and RAS integrated with BB and AA during specific periods. They appear to be associated with the most favourable prognosis.

\subsection{A posteriori analysis of the extreme group compared to the reference} 
Clusters derived from the combined RAS, BB, and AA sequences demonstrate notable significance in the proposed survival models. The survival analysis reveals the influence of the administered therapy type on the patient's probability of survival.
Looking at Fig.~\ref{fig:HR}, it becomes evident that being a member of cluster 2 significantly enhances the odds of survival compared to cluster 1. We present a descriptive study to delve deeper into patient differences between these two clusters. Table~\ref{tab:cluster_comp} illustrates the variable distributions between the two groups, facilitating a post hoc analysis of patient characteristics. The Kruskal-Wallis test compares categorical variable distributions, while the Wilcoxon test is used for continuous ones. Table~\ref{tab:cluster_comp} shows that cluster 2 predominantly comprises younger male patients with better clinical conditions since low values of MCS characterize it. As expected, since belonging to cluster 2 is a protective factor in the survival analysis, the proportion of dead patients is small and corresponds to 450 patients (14.7\%).

\begin{table}[t]
\caption{Comparison between reference (cluster 1) and cluster 2 patients' characteristics.}
\centering
\begin{tabular}{lllll}
\hline
 & \textbf{Value} & \textbf{Cluster 1} & \textbf{Cluster 2} & \textbf{p-value} \\ 
\noalign{\smallskip}\hline\noalign{\smallskip}
\textbf{N° of patients} &  & 5,557 & 3,059 &  \\ \hline
\textbf{Age} & mean (sd) & 76.5 (12.3) & 69.4 (11.4) & $<2\cdot 10^{-16}$ \\ \hline
\textbf{Sex} & Male (\%) & 2,618 (47.1) & 1,818 (59.4) & $<2\cdot 10^{-16}$\\
\textbf{} & Female (\%) & 2,939 (52.9) & 1,241 (40.6) &  \\ \hline
\textbf{Death} & 0 (\%) & 3,425 (61.6) & 2,609 (85.3) & $<2\cdot 10^{-16}$ \\
\textbf{} & 1 (\%) & 2,132 (38.4) & 450 (14.7) &  \\ \hline
\textbf{Days in hospital} & mean (sd) & 19.7 (30.4) & 11.2 (19.1) & $<2\cdot 10^{-16}$ \\ \hline
\textbf{Total procedures} & 0 (\%) & 5,157 (92.8) & 2,519 (82.4) & $<2\cdot 10^{-16}$ \\
\textbf{} & 1 (\%) & 378 (6.8) & 527 (17.2) &  \\
\textbf{} & $\geq$ 2 (\%) & 22 (0.4) & 13 (0.4) &  \\ \hline
\textbf{MCS} & Low (\%) & 2,130 (38.5) & 1,437 (47) &  $<2\cdot 10^{-16}$\\
 & Intermediate (\%) & 2,772 (49.9) & 1,468 (48) &  \\
 & High (\%) & 655 (11.8) & 154 (5) &  \\ 
 \noalign{\smallskip}\hline\noalign{\smallskip}
\end{tabular}
\label{tab:cluster_comp}
\end{table}

\section{Discussion and conclusions}
This research represents a methodological exploration into the impact of different drug-utilization patterns on the survival of heart failure patients, leveraging administrative databases from the Lombardy region. 
The work included two main parts, focusing first on extrapolating information from the longitudinal data in the administrative database to obtain workable and complete drug histories of the patients. The necessity to describe complex patterns of care over time while preserving the complexity and variability of this information drives the application of state-sequence analysis techniques. The state-sequence analysis allows the description of drug-utilization patterns for each patient through a mathematical object, which provides tools for assessing and associating the primary endpoint of interest. 
Compared to commonly used baseline measures, which omit some time-dependent information, this technique yields more realistic and valuable results since it considers the entire evolution of each patient's clinical path. In particular, it allows observing the specific weeks and duration the therapies cover the patient after the first hospitalization. State-sequence analysis and clustering techniques allowed us to identify latent patterns describing different drug assumption behaviors. The construction of the combined sequence that, for each week of observation, describes the prescription of the three main pharmaceutical classes, RAS, BB, and AA, simultaneously makes it possible to study the patient's behavior for the different drugs. This procedure, i.e., the analysis of the polytherapy taken by the patient, is innovative in this type of problem. Using a single patient descriptor (the sequence), the SSA makes it possible to study the prescription of different kinds of drug and their combination in each week of observation, i.e., different polytherapies. Indeed, the clustering of those sequences reveals eight distinct behaviors, spotting multi-drug combination patterns.
In the second part of this work, we study the effect of belonging to different groups that emerged by the SSA-based cluster analysis on patients' survival probability. Patient characterizations obtained from sequence clustering revealed significant insights in prediction. Five out of seven clusters identifying different path therapies showed significant differences from the reference cluster that identified the lack of therapy intake for almost the entire observation period. This shows that belonging to a group representing the prescription of at least one of the possible therapies, excluding the monotherapy with anti-aldosterone agents, is a protective factor on the survival probability. Notably, even undergoing one therapy for half the observation time benefits survival. From the survival analysis, cluster 2, which identifies patients with regular intakes of RAS and BB two-drug combinations, shows the highest survival probability compared to the reference cluster. Their patients' characteristics are carefully compared, and it emerges that the latter group has a more severe clinical condition due to many comorbidities. Looking at this characteristic and adding that most are female with an average age of 76.5 years, the literature suggests that this cluster identifies patients with heart failure with a preserved ejection
fraction~\cite{cite_15}. It can be concluded that even if we do not know the severity of the HF, taking drugs for an extended period after the first hospitalization seems essential to improve the prognosis of heart failure.

The SSA, therefore, allows a change of perspective in the analysis of the prescriptions, moving from a transversal and syntactical approach to a holistic one that exploits the information available through the application of statistical tools, slightly more complex than traditional methods~\cite{cite_32}.
This work represents an important step in evaluating HF patients' drug-based path and paves the way for future developments. Latent Markov Models can be applied to study hidden patterns in this data to investigate the evolution of an individual characteristic that is not directly observable~\cite{cite_33}. Future expansions regard the application of these techniques to the case where multiple sources of information are described through SSA, allowing us to describe individual trajectories on several dimensions simultaneously~\cite{cite_30, cite_34}.

Our study also has some limitations. Firstly, due to the administrative nature of our dataset, we had to deal with some restrictions: (i) the Lombardy regional dataset contains restricted clinical information (it is sufficient to consider that the information on comorbidities is obtained only from the hospitalizations of patients, thus identifying only severe adverse events); (ii) the clinical data characterizing heart failures (e.g., blood pressure, the ejection fraction), information about contraindications/intolerance to drugs, or the results of diagnostic tests are not included; (iii) only information about drug purchases are registered, but we do not have information on actual drug dose prescriptions provided by doctors and also they do not coincide with the effective administration of the therapy, leading to a potential wrong calculation of the drug coverage. For this reason, exposure misclassification could affect these results since the duration of dispensed drugs is calculated according to the defined daily dose metric; (iv) this dataset contains information only on patients treated by the public health service, losing the information on patients followed privately, even if generally, the latter represents a minority, especially regarding the purchase of drugs, which is guaranteed free of charge by the Italian National Health Service (NHS).
Secondly, the study's observation period was limited to a single year to guarantee a clear presentation of the various sequences within and between clusters. Further investigation should apply sequence clustering approaches over a more extended period to assess the long-term impacts. Thirdly, the less frequent paths are left out of the display due to the typical sequences' extraction, which was done to highlight the clusters' key features. This implies that part of the collected variability in the patient care paths can be lost upon clustering. Lastly, the analysis's findings could change based on the arbitrary selection of the clustering technique and dissimilarity metrics~\cite{cite_10}. No findings indicate that one approach is better than another. Sensitivity analysis and subject expertise are therefore necessary.  

In conclusion, SSA has provided impressive results in studying drug-utilization patterns of heart failure patients. However, more specific and updated data can lead to even more effective results that can support healthcare specialists in evaluating the pathways provided to patients and the National Health Service in allocating the most appropriate resources. Additionally, because there are so many possibilities for defining states, temporal granularity, and spacing metrics, SSA is a flexible method that may be used in various epidemiological scenarios.

%\bibliographystyle{unsrtnat}
%\bibliography{references}  %%% Uncomment this line and comment out the ``thebibliography'' section below to use the external .bib file (using bibtex) .
%\begin{thebibliography}{99.}

%\begin{thebibliography}{.99}

\newpage
%%%%%%%%%%%%%%%%%%%%%%%% referenc.tex %%%%%%%%%%%%%%%%%%%%%
% sample references
% 
% Use this file as a template for your own input.
%
%%%%%%%%%%%%%%%%%%%%%%%% Springer%%%%%%%%%%%%%%%%%%%%%%%%%%
%
% BibTeX users please use
% \bibliographystyle{}
% \bibliography{}

\begin{thebibliography}{99.}%
% MY BIBLIOGRAPHY

%1) Introduction
\bibitem{cite_intro_1}Dyck GJB, Raj P, Zieroth S, Dyck JRB, Ezekowitz JA. The Effects of Resveratrol in Patients with Cardiovascular Disease and Heart Failure: A Narrative Review. International Journal of Molecular Sciences, \textbf{20}(4):904 (2019)

\bibitem{cite_intro_2}Roger, V. L. (2021). Epidemiology of Heart Failure. Circulation Research, \textbf{128}(10), 1421–1434 (2021)

\bibitem{cite_1}Rossignol, P., Hernandez, A. F., Solomon, S. D., Zannad, F.: Heart failure drug treatment. Lancet, \textbf{393}, 1034-1044 (2019)

\bibitem{cite_15} McDonagh, T. A., Metra, M., Adamo, M., Gardner, R. S., Baumbach, A., Böhm, M., Burri, H., Butler, J., Čelutkienė, J., Chioncel, O., Cleland, J. G. F., Coats, A. J. S., Crespo-Leiro, M. G., Farmakis, D., Gilard, M., Heymans, S., Hoes, A. W., Jaarsma, T., Jankowska, E. A., Lainscak, M., … 2021 ESC Guidelines for the diagnosis and treatment of acute and chronic heart failure. Eur. Heart J. \textbf{42}, 3599–3726 (2021)

\bibitem{cite_27}Swedberg, K., Cleland, J., Dargie, H., Drexler, H., Follath, F., Komajda, M., Tavazzi, L., Smiseth, O. A., Gavazzi, A., Haverich, A., Hoes, A., Jaarsma, T., Korewicki, J., Lévy, S., Linde, C., Lopez-Sendon, J. L., Nieminen, M. S., Piérard, L., Remme, W. J., \& Task Force for the Diagnosis and Treatment of Chronic Heart Failure of the European Society of Cardiology. Guidelines for the diagnosis and treatment of chronic heart failure: executive summary (update 2005): The Task Force for the Diagnosis and Treatment of Chronic Heart Failure of the European Society of Cardiology. European heart journal, 26(11), 1115–1140 (2005) 

\bibitem{cite_13} McMurray, J. J., Adamopoulos, S., Anker, S. D., Auricchio, A., Böhm, M., Dickstein, K., Falk, V., Filippatos, G., Fonseca, C., Gomez-Sanchez, M. A., Jaarsma, T., Køber, L., Lip, G. Y., Maggioni, A. P., Parkhomenko, A., Pieske, B. M., Popescu, B. A., Rønnevik, P. K., Rutten, F. H., Schwitter, J., … ESC Guidelines for the diagnosis and treatment of acute and chronic heart failure 2012: The Task Force for the Diagnosis and Treatment of Acute and Chronic Heart Failure 2012 of the European Society of Cardiology. Developed in collaboration with the Heart Failure Association (HFA) of the ESC. Eur. Heart J. \textbf{33}, 1787–1847 (2012)

\bibitem{cite_14} Probstfield, J.L., O'Brien, K.D. Progression of Cardiovascular Damage: The Role of Renin–Angiotensin System Blockade. Am. J. Cardiol. \textbf{105}, 10A-20A (2012)


%Mortality/hospitalization rates
\bibitem{cite_16}Lan, T., Liao, Y. H., Zhang, J., Yang, Z. P., Xu, G. S., Zhu, L., Fan, D. M. Mortality and Readmission Rates After Heart Failure: A Systematic Review and Meta-Analysis. Ther. Clin. Risk. Manag. \textbf{19}, 1307–1320 (2021)

\bibitem{cite_17}Corrao, G., Ghirardi, A., Ibrahim, B., Merlino, L., Maggioni, A. P. Short- and long-term mortality and hospital readmissions among patients with new hospitalization for heart failure: A population-based investigation from Italy. Int. J. Cardiol. \textbf{181}, 81–87 (2015) 

%Link with non-adherence
\bibitem{cite_18} Fitzgerald, A.A., Powers, J.D., Ho, P.M., Maddox, T.M., Peterson, P.N., Allen, L.A., et al. Impact of Medication Nonadherence on Hospitalizations and Mortality in Heart Failure. JCF \text{17}, 664–669 (2011)

\bibitem{cite_19}Ruppar, T., Cooper, P., Mehr, D., Delgado, J. \& Jacqueline M. DJ. Medication Adherence Interventions Improve Heart Failure Mortality and Readmission Rates: Systematic Review and Meta‐Analysis of Controlled Trials. JAHA \textbf{5}, (2016)

\bibitem{cite_20}Oosterom-Calo, R., van Ballegooijen, A.J., Terwee, C.B. et al. Determinants of adherence to heart failure medication: a systematic literature review. Heart Fail Rev \textbf{18}, 409–427 (2013).

\bibitem{cite_21}Aggarwal, B., Pender, A., Mosca, L., \& Mochari-Greenberger, H. Factors associated with medication adherence among heart failure patients and their caregivers. J. Nurs. Educ. Pract. \textbf{5}(3), 22–27 (2015)

\bibitem{cite_3}Wu, J., Moser, D. K. and Lennie, T. A., Burkhart, P. V.: Medication adherence in patients who have heart failure: a review of the literature. Nursing Clinics Of North America. \textbf{43}, 133-53; vii–viii (2008)

\bibitem{cite_28} Kim, E.S., Youn, J.C., Baek, S.H. Update on the pharmacotherapy of heart failure with reduced ejection fraction. Cardiovascular Prevention and Pharmacotherapy \textbf{2}, 113–133 (2020)

\bibitem{cite_22} Scalvini, S., Bernocchi, P., Villa, S., Paganoni, A.M., La Rovere, M.T., Frigerio, M. Treatment prescription, adherence, and persistence after the first hospitalization for heart failure: A population-based retrospective study on 100785 patients. Int. J. Cardiol. \textbf{330}, 106–111 (2021)

\bibitem{cite_23} Spreafico, M., Gasperoni, F., Barbati, G., Ieva, F., Scagnetto, A., Zanier, L., Iorio, A., et al. Adherence to Disease-Modifying Therapy in Patients Hospitalized for HF: Findings from a Community-Based Study. Am. J. Cardiovasc. Drugs \textbf{20}, 179–190 (2020)

\bibitem{cite_24} Di Martino, M., Alagna, M., Lallo, A., Gilmore, K.J., Francesconi, P., Profili, F., et al. Chronic polytherapy after myocardial infarction: the trade-off between hospital and community-based providers in determining adherence to medication. BMC Cardiovasc. Disord. \textbf{21}, 180 (2021) 

%SSA
\bibitem{cite_25} Savaré, L., Ieva, F., Corrao, G. et al. Capturing the variety of clinical pathways in patients with schizophrenic disorders through state sequences analysis. BMC Med Res Methodol \textbf{23}, 174 (2023).

\bibitem{cite_26} Roux, J., Grimaud, O., Leray, E. Use of state sequence analysis for care pathway analysis: The example of multiple sclerosis. Statistical methods in medical research, \textbf{28}(6), 1651–1663 (2019)

%Administrative data
\bibitem{cite_29}Mathew, S., Peat, G., Parry, E., Sokhal, B. S., Yu, D. Applying Sequence Analysis to Uncover ‘Real-World’ Clinical Pathways from Routinely Collected Data: A Systematic Review. Journal of Clinical Epidemiology, 111226 (2023) 

\bibitem{cite_7}Timofte, D., Pantea Stoian, A., Razvan, H., Diaconu, C., Bulgaru-Iliescu, D., Balan, G., Ciuntu, B., Neagoe, R.: A Review on the Advantages and Disadvantages of Using Administrative Data in Surgery Outcome Studies. The Journal Of Surgery. \textbf{14} (2018)

%2) Cohort selection
\bibitem{cite_8}Rea, F., Savaré, L., Franchi, M., Corrao, G., Mancia, G.: Adherence to Treatment by Initial Antihypertensive Mono and Combination Therapies. American Journal Of Hypertension. \textbf{34}, 1083-1091 (2021)

\bibitem{cite_21_LMM} Centers for Disease Control and Prevention. International Classification of Diseases, Ninth Revision, Clinical Modification (ICD-9-CM). \url{https://www.cdc.gov/nchs/icd/icd9cm.htm. Cited 15 December 2023} 
%(12 July 2023)

\bibitem{cite_22_LMM} World Health Organization. Anatomical Therapeutic Chemical (ATC) Classification. \url{https://www.who.int/tools/atc-ddd-toolkit/atc-classification. Cited 15 December 2023} %(12 July 2023)

\bibitem{cite_23_LMM} World Health Organization. Adherence to long-term therapies: Evidence for action. WHO Library Cataloguing-in-Publication Data. (2003).

\bibitem{cite_9}Corrao, G., Rea, F., Di Martino, M., De Palma, R., Scondotto, S., Fusco, D., Lallo, A., Belotti, L. M. B., Ferrante, M., Pollina Addario, S., Merlino, L., Mancia, G., Carle, F. Developing and validating a novel multisource comorbidity score from administrative data: a large population-based cohort study from Italy. BMJ Open. \textbf{7} (2017)


%3) Methodologies
%3.1) SEQUENCE DESCRIPTION
\bibitem{cite_10}Studer, M., Ritschard, G. What Matters in Differences Between Life Trajectories: A Comparative Review of Sequence Dissimilarity Measures. Journal of the Royal Statistical Society Series A: Statistics in Society, \textbf{179}(2), 481–511 (2015)

\bibitem{cite_11}Gabadinho, A., Ritschard, G., M{\"u}ller, N. S., Studer, M.: Analyzing and visualizing state sequences in R with TraMineR. J. Stat. Softw. \textbf{40} (2011)

\bibitem{cite_30}Emery, K., Berchtold, A.  Comparison of two approaches in multichannel sequence analysis using the Swiss Household Panel. Longitudinal and life course studies: international journal, \textbf{14}(4), 592–623 (2022) 

\bibitem{cite_13_TESI} Studer, M., Ritschard, G. Sequence Analysis and Related Approaches. Springer Cham, (2018)

\bibitem{cite_44_TESI} Han, Y., Liefbroer, A. C., Elzinga, C. H.  Comparing methods of classifying life courses: sequence analysis and latent class analysis. Longitudinal and Life Course Studies, \textbf{8}(4), 319-341 (2017)

\bibitem{cite_12}Lesnard, L.: Setting Cost in Optimal Matching to Uncover Contemporaneous Socio-Temporal Patterns. Sociological Methods \& Research. \textbf{38}, 389-419 (2010)

%clustering seq
\bibitem{cite_PBC}Kornbrot, D. Point Biserial Correlation. In Encyclopedia of Statistics in Behavioral Science (eds B.S. Everitt and D.C. Howell) (2005)

\bibitem{cite_HC}Hubert, L. J., Levin, J. R. A general statistical framework for assessing categorical clustering in free recall. Psychological Bulletin, \textbf{83}(6), 1072–1080 (1976)

\bibitem{cite_ASW}Batool, F., Hennig, C. Clustering with the Average Silhouette Width. Computational Statistics \& Data Analysis, \textbf{158}, 107190 (2021)

\bibitem{cite_31}Studer, M. WeightedCluster Library Manual: A practical guide to creating typologies of trajectories in the social sciences with R. LIVES Working Papers, \textbf{24} (2013)


%4) Results
\bibitem{cite_31bis} Suissa, S. Immortal time bias in pharmaco-epidemiology. Am J Epidemiol. 167(4):492-9 (2008)

Studer, M. WeightedCluster Library Manual: A practical guide to creating typologies of trajectories in the social sciences with R. LIVES Working Papers, \textbf{24} (2013)


%5) Discussion
\bibitem{cite_32}Vanoli, J., Nava, C. R., Airoldi, C., Ucciero, A., Salvi, V., Barone-Adesi, F. Use of State Sequence Analysis in Pharmacoepidemiology: A Tutorial. International journal of environmental research and public health, 18(24), 13398 (2021)

\bibitem{cite_33} Fontana, N., Savaré, L., Rea, F., Angelantonio, E. D., Ieva, F. Long-term adherence to polytherapy in heart failure patients: a novel approach emphasising the importance of secondary prevention. arXiv [Stat.AP] (2023)


\bibitem{cite_34}Ritschard, G., Liao, T. F., Struffolino, E. Strategies for Multidomain Sequence Analysis in Social Research. Sociological Methodology, \textbf{53}(2), 288-322 (2023)



%% TEMPLATE
% Use the following syntax and markup for your references if 
% the subject of your book is from the field 
% "Mathematics, Physics, Statistics, Computer Science"
%
% Contribution 
%\bibitem{science-contrib} Broy, M.: Software engineering --- from auxiliary to key technologies. In: Broy, M., Dener, E. (eds.) Software Pioneers, pp. 10-13. Springer, Heidelberg (2002)
%
% Online Document
%\bibitem{science-online} Dod, J.: Effective substances. In: The Dictionary of Substances and Their Effects. Royal Society of Chemistry (1999) Available via DIALOG. \\
%\url{http://www.rsc.org/dose/title of subordinate document. Cited 15 Jan 1999}
%
% Monograph
%\bibitem{science-mono} Geddes, K.O., Czapor, S.R., Labahn, G.: Algorithms for Computer Algebra. Kluwer, Boston (1992) 
%
% Journal article
%\bibitem{science-journal} Hamburger, C.: Quasimonotonicity, regularity and duality for nonlinear systems of partial differential equations. Ann. Mat. Pura. Appl. \textbf{169}, 321--354 (1995)
%
% Journal article by DOI
%\bibitem{science-DOI} Slifka, M.K., Whitton, J.L.: Clinical implications of dysregulated cytokine production. J. Mol. Med. (2000) doi: 10.1007/s001090000086 
%
%\bigskip

% Use the following (APS) syntax and markup for your references if 
% the subject of your book is from the field 
% "Mathematics, Physics, Statistics, Computer Science"
%
% Online Document
%\bibitem{phys-online} J. Dod, in \textit{The Dictionary of Substances and Their Effects}, Royal Society of Chemistry. (Available via DIALOG, 1999), 
%\url{http://www.rsc.org/dose/title of subordinate document. Cited 15 Jan 1999}
%
% Monograph
%\bibitem{phys-mono} H. Ibach, H. L\"uth, \textit{Solid-State Physics}, 2nd edn. (Springer, New York, 1996), pp. 45-56 
%
% Journal article
%\bibitem{phys-journal} S. Preuss, A. Demchuk Jr., M. Stuke, Appl. Phys. A \textbf{61}
%
% Journal article by DOI
%\bibitem{phys-DOI} M.K. Slifka, J.L. Whitton, J. Mol. Med., doi: 10.1007/s001090000086
%
% Contribution 
%\bibitem{phys-contrib} S.E. Smith, in \textit{Neuromuscular Junction}, ed. by E. Zaimis. Handbook of Experimental Pharmacology, vol 42 (Springer, Heidelberg, 1976), p. 593

\end{thebibliography}
%
%\biblstarthook{

%\end{thebibliography}

\end{document}